\documentclass[sigconf]{acmart}

\usepackage{xcolor}
\usepackage{color}
\usepackage{graphicx}
\usepackage{subfigure}
\usepackage{multirow}
\usepackage{array}
\usepackage{xspace}
\usepackage{enumitem} 
\usepackage{fontawesome5}

\usepackage{tcolorbox}

\AtBeginDocument{%
  \providecommand\BibTeX{{%
    \normalfont B\kern-0.5em{\scshape i\kern-0.25em b}\kern-0.8em\TeX}}}

\setcopyright{none}

\acmConference[Conference acronym 'XX]{Make sure to enter the correct
  conference title from your rights confirmation emai}{June 03--05,
  2018}{Woodstock, NY}

\begin{document}

\title{Teaching Software Ethics to Future Software Engineers }








 \author{Aastha Pant \hspace{0.2cm}    Simone V. Spiegler \hspace{0.2cm}      Rashina Hoda}        
 \author{Jeremy Yoon \hspace{0.2cm}     Nabeeb Yusuf \hspace{0.2cm}   Tian Er \hspace{0.2cm}        Shenyi Hu}

\email{{aastha.pant, simone.spiegler, rashina.hoda}@monash.edu} 
\email{{jyoo0005, nyus0004, terr0001, shuu0020}@student.monash.edu}
 
\affiliation{%
 \institution{Faculty of Information Technology, Monash University}
    \city{Melbourne}
  \state{Victoria}
  \country{Australia}}

\renewcommand{\shortauthors}{Pant, et al.}

\begin{abstract}
The importance of teaching software ethics to software engineering (SE) students is more critical now than ever before as software-related ethical issues continue to impact society at an alarming rate. Traditional classroom methods, vignettes, role-play games, and quizzes have been employed over the years to teach SE students about software ethics. Recognising the significance of incorporating software ethics knowledge in SE education and the continued need for more efforts in the area of the teaching and learning of SE ethics, we developed an interactive, scenario-based \emph{Software Ethics Quiz}. Our goal was to teach SE students about ethics in a comprehensive, open, and engaging manner through a combined approach of an online lecture followed by an interactive workshop with the quiz and a debriefing session. The anonymous quiz responses collected showed promising results regarding the engagement and efficacy of the lecture and quiz, with a slightly better rating for the interactive quiz. The voluntary student feedback collected suggested that a majority of the participants found the debrief discussion on the quiz scenarios to be very beneficial for learning about software ethics. In this experience report, we share our experiences, related educational resources including the quiz, and recommendations from lessons learned with the wider education community to keep driving this critical topic forward.
\end{abstract}

\ccsdesc[500]{Software engineering~Software ethics}

\keywords{ethics, education, quiz, engagement, software engineering}

\maketitle

\section{Introduction} \label{sec:introduction}

The importance of teaching ethics to software engineering (SE) students has been acknowledged over the years \cite{ryan2020we,CCSE}. It has never been more pertinent than in current times where software is increasingly becoming a necessity to life's essential facets, including how we receive news, connect with loved ones, and access healthcare, reaching and affecting millions of users worldwide. The increasingly ubiquitous nature of software across domains has led to a significant rise in ethical concerns related to its use. This issue is particularly pronounced in the context of artificial intelligence-based software \cite{al2023ab,Amazon}. In recent times, various ethical concerns have surfaced, exemplified by incidents such as the notorious data breach involving Cambridge Analytica, where the private data of 87 million users was accessed and subsequently leaked to the public \cite{Cambridge}. 
Instances of plane crashes attributed to software bugs, leading to four fatalities \cite{Airbus}. The Annual Data Breach Investigations Report highlights carelessness and errors as among the top three reasons behind data breach incidents, underscoring the significance of these issues \cite{bassett2021data}. Similarly, in 2021, Toshiba Tech Group, a technology company, experienced a cyber-attack conducted by a hacker group named DarkSide. In response, Toshiba committed to enhancing its security measures to safeguard customer data \cite{Toshiba}. In 2021, a significant software flaw was found in Log4j, which enabled attackers to remotely execute code on susceptible servers through specially crafted requests \cite{Log4}. 
The growing list of such incidents underscores the critical need to prepare future software engineers by improving their awareness and understanding of SE ethics.

A variety of approaches have emerged over the years to support the teaching and learning of SE ethics. These vary from traditional classroom settings to vignettes and role-play games, all in a bid to raise ethics awareness among students with varying degrees of success \cite{bairaktarova2015engineering}. Traditional classroom-based approaches are acknowledged to be lacking in student engagement, leading to poor effectiveness \cite{lane2015new}. Other approaches such as vignettes and role-play games are seen to be more promising in improving engagement. However, the disparity between participants' abstract descriptions of their actions and the actual occurrences in reality have been acknowledged as challenges \cite{skilling2020using}. At the same time, executing a role-play involves thorough planning and consideration, coupled with continuous efforts to engage and prepare the class or students assigned roles throughout the session \cite{blatner2009role}. Role-play also requires acting, which may not come easy to many instructors.

Along the same lines, quizzes have been used for SE ethics education in the past \cite{ryan2020we} and for AI ethics training more recently \cite{teo2023would}. While not without limitations, quizzes are seen to offer a valuable reference for later lectures on law, ethics, and morals, especially in grounding discussions on codes of practice and professional ethics for the students \cite{ryan2020we}. Moreover, quizzes contribute to self-assessment of ethical knowledge, pinpointing areas of improvement and enhancing overall ethical practices in everyday development scenarios \cite{teo2023would}.

With an aim to motivate similar experiential learning, we explored the use of a scenario-based Software Ethics Quiz in a higher education software engineering classroom. Overall, we wanted to explore the experience of introducing ethics to SE students through an online lecture and associated resources, followed by an interactive in-person quiz and debrief session. In particular, we wanted to explore the following research questions:

\begin{itemize}
    \item \textbf{RQ1}: To what extent do software engineering students make ethical decisions when faced with hypothetical scenarios?
    \item \textbf{RQ2}: How helpful is an online lecture in improving students' awareness and understanding of ethics? How helpful is an interactive in-person quiz for the same?
    \item \textbf{RQ3}: How engaging is an online lecture to learn about ethics in software engineering? How engaging is an interactive in-person quiz for the same?
    \item \textbf{RQ4}: How difficult is it to find the ideal answers to the quiz questions?
    \item \textbf{RQ5}: How helpful is a debrief discussion on the quiz’s scenarios and student responses to learn about ethics in software engineering?
\end{itemize}

Motivated by this aim and questions, while remaining open to other insights, we designed the quiz to include three hypothetical scenarios inspired by ethical dilemmas brought to the fore in real-world cases of ethical breaches. Each scenario was accompanied by three questions with four answer choices. Each choice is related to one potential action when facing this scenario. Since ethics is hardly ever black or white, we designed these answer options to range from ideal, desirable, to bearable, and wrong. The quiz was used in a second-year Bachelor of Software Engineering class focused on the SE process, taken by approximately 400 students including students from SE, computer science, and others. Once students had taken the quiz individually, the instructor conducted a debriefing session where they went through the scenarios and options with the class as a whole, eliciting active student voice, debate, and discussions. Finally, students were requested to share their experiences through a voluntary feedback survey. We analysed our quantitative data using descriptive statistics. Details of the quiz design, scenarios and answers, results of the quiz, and findings from the feedback survey are shared in the upcoming sections of this experience report.

This study's primary contributions lie in the development of a \emph{Software Ethics Quiz} designed to educate SE students about ethics, which we share with the wider SE education community for further development, use, and improvements. Additionally, we share our experiences in teaching ethics and discuss the lessons learned in the process. These educational resources and insights can prove beneficial for SE educators aiming to enhance awareness of software ethics among their students.

\section{Background and Related Work}\label{sec:background}

While there is no universally accepted definition of ethics, in this work, we use the following two definitions of ethics, ``\textit{The moral principles that govern the behaviors or activities of a person or a group of people}'' \cite{nalini2020hitchhiker} and ``\textit{The way an individual behaves and the values they hold}'' \cite{iacovino2002ethical}. 

In the present era, nearly every sector relies on software, and its extensive application in domains such as information technology, healthcare, transport, banking, and others has increased the ethical concerns associated with its use. For example, there have been several software-related ethical incidents in recent years, such as the Hawaii False Missile Alert \cite{Hawaii}, a buggy vaccine scheduling system in New Jersey hospitals \cite{Buggy}, Airbus’ buggy software causing the crash of an A400M plane and four deaths \cite{Airbus}, and many data breaches \cite{TMobile,TMX} that have highlighted the importance of ethics during software development. The increasing prevalence of Digital Addiction, as discussed by \citet{ali2015emerging}, and the existence of discriminatory software, as outlined by \citet{galhotra2017fairness}, clearly illustrate the detrimental impacts of the lack of ethical considerations in software on modern society.

Software engineering code of ethics and ethical guidelines have been developed worldwide by multiple organisations such as the Association for Computing Machinery (ACM) \cite{ACM}, the Australian Computer Society \cite{ACS}, the British Computer Society (BCS) \cite{BCS}, the IEEE Computer Society (IEEE-CS) \cite{IEEECS}, the New Zealand Computer Society \cite{ITP} and many more to ensure the development of ethical software \cite{gotterbarn2001software}. These codes and guidelines serve to emphasise the importance of including ethics as part of SE education. For example, in the guiding principles of the Software Engineering segment within the Computing Curricula, presently in the process of further development under the IEEE Computer Society (IEEE-CS) and the Association for Computing Machinery (ACM), there is a clear acknowledgment that: ``\emph{The education of all Software Engineering students must include student experiences with the professional practice of Software Engineering. The Professional Practice of Software Engineering encompasses a wide range of issues and activities including problem-solving, management, ethical and legal concerns...}'' \cite{CCSE}. Along with developing ethical guidelines, emphasis has been placed on raising awareness of software ethics among SE professionals.

Likewise, various studies have emphasized the importance of teaching ethics to the students in the Software Engineering Curriculum \cite{towell2004further,ryan2020we}. \citet{ryan2020we} underscores that software students must be taught professional ethics within SE curricula as the industry may not prioritise this aspect. 

However, teaching SE students about these codes of ethics in a way that is both engaging and effective is a non-trivial pursuit for SE educators. This is because a lecture on SE ethics can easily become a boring list of do's and don'ts. For example, research shows that requesting participants to contemplate ethical vignettes based on the ACM Code of Ethics has no influence on ethical decision-making \cite{mcnamara2018does}. Various quizzes such as \emph{Moral Machine} \cite{awad2018moral} and \emph{Interactive AI Ethics Quiz} \cite{teo2023would} have been developed with the aim of raising awareness of ethics among the public and software/AI professionals respectively. 
 
The topic of educating SE students on ethics continues to be explored and studied. Several efforts have been undertaken to increase students' awareness of software ethics through traditional education and training, but much remains to be done in this area. For instance, \citet{bairaktarova2015engineering} reported that the students who had taken the ethics class did not perform well in an ethical test. \citet{berry2010ethics} express a similar view through their test, examining students' ethical behavior both before and after ethics instruction. Despite participating in debates and in-class discussions on the topic, there was no observable improvement in the students' \textit{``ethicality"}. 
Such outcomes may be attributed to the fact that traditional classroom-based and didactic approaches frequently lack engagement and may not be effective \cite{lane2015new}. \citet{towell2003teaching} carried out a survey indicating that the primary methods employed by a majority of their participants (educators) 
for teaching ethics to SE students were discussing personal experiences (58\%) and codes of ethics (58.30\%). Additional approaches included engaging in discussions based on case studies and assigned readings and research papers. 

Taking into account these factors, coupled with the highlighted importance of, and varying levels of success in, educating SE students on ethics in the literature, we decided to introduce SE students to software ethics in an interactive way and capture the lessons related to the experience. To accomplish this, we developed a scenario-based Software Ethics Quiz and evaluated it with the cohort of SE students as described next.

\section{Method}\label{sec:method}

\subsection{Software Ethics Quiz -- Design and Development}
\begin{figure*}[htbp]
    \centering
    \includegraphics [scale=0.50] {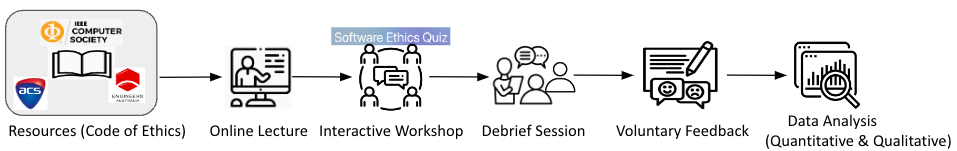}
    \caption{Teaching design used in our study}
    \label{fig:Method}
\end{figure*}

Our objective was to develop a scenario-based Software Ethics Quiz to enable SE students to actively think and interactively learn about software ethics. Figure \ref{fig:Method} shows the teaching design we used in our study. 
\subsubsection{Quiz Content}

\begin{table*}[p] 
\centering
\scriptsize
\caption{Mapping answers to questions and linking them to the applicable code of ethics}
\label{tab:MappingTable}
\begin{tabular} {>{\raggedright\arraybackslash}p{4cm}>{\raggedright\arraybackslash}p{4.4cm}>{\raggedright\arraybackslash}p{3.8cm}>{\raggedright\arraybackslash}p{1.5cm}>{\raggedright\arraybackslash}p{1.3cm}} 

\hline

Scenario & Question & Answer & Evaluation & Code of Ethics\\
\hline

Optus became a victim of a cyberattack that leaked up to 10 million Australians’ personal data including phone numbers and email addresses. You are a software developer in a company that suffered from a similar security breach on user’s data.  & The attackers asked for a ransom, or they would leak the data. What would be the best course of action in this scenario?
 & Threaten attackers with legal actions & Wrong & The Primacy of the Public Interest (ACS), Public (ACM)  \\

&  & Ignore demand for ransom & Not desirable & \\
& & Pay ransom and secure leaked data & Bearable & \\
& & Seek help from authorities & Ideal & \\
\cline{2-5}

& Your customers’ data has been compromised. What should you convey to your customers? & Attempt to cover up the situation. Don't tell the public to avoid backlash & Wrong & Honesty (ACS), Demonstrate Integrity (EA) \\

& & Tell customers there was a breach but not what was leaked & Not desirable & \\
& & Tell customers what was leaked and reassure data recovery & Bearable & \\
& & Tell customers what was leaked and what actions they can take & Ideal & \\

\cline{2-5}

& You notice you made a mistake that caused a security vulnerability when implementing the cybersecurity measures, but nobody else noticed and your work passed the review and the update to the system has been deployed. What would you do? & Keep quiet and enjoy your team’s success at another software update & Wrong & Honesty (ACS), The Primacy of the Public Interest (ACS), Public (ACM), Demonstrate Integrity (EA) \\

& & Report the issue but don't recall the update & Not desirable & \\
& & Recall the update and identify the code reviewer who missed the mistake & Bearable & \\
& & Report the issue and try to recall the update & Ideal & \\

\hline

Facebook and Twitter use sophisticated algorithms to determine the exact combination of notifications, icons, and eye-catching features to keep users hooked. You are part of a company developing a social media application called Convey, similar to Facebook. To drive profits, your company has gradually implemented more features into Convey to drive user engagement. & Features aimed at maximising user retention are known to be addictive and negatively affect user's mental health. Who should act first to reduce the harm? & Stakeholders (investor) of the company & Wrong & Promote Sustainability (EA), The Primacy of the Public Interest (ACS), Public (ACM) \\

& & Government and regulatory bodies & Not desirable & \\
& & Developers of the application & Bearable & \\
& & CEO and managers of the company & Ideal & \\

\cline{2-5}

& In an attempt to boost user engagement, echo chambers have been created on Convey. This has led to the propagation of misinformation and has created a larger divide in the user base. What should Convey’s first course of action be? & Allow politicians to invest and capitalise on the polarisation & Wrong & Promote sustainability (EA) \\

& & Wait for regulations to change for all social media platforms & Not desirable & \\
& & Campaign for change in regulations for all social media platforms & Bearable  & \\
& & Remove algorithms that contribute to these echo chambers & Ideal & \\

\cline{2-5}
& What should Convey do to become a revolutionary social media platform? & Provide more personalised content than Facebook and Twitter & Wrong & Exercise leadership (EA) \\

& & Sanitise content for under aged users & Not desirable & \\
& & Personalise content without polarising user base & Bearable  & \\
& & Introduce diversity in the content recommendation algorithm & Ideal & \\

\hline

Dona is an entry-level software engineer at SDX Alliance and is assigned to develop a new hard disk controller. They reuse code from their previous job at a start-up without informing anyone. During the review process, it was discovered that the code belonged to their former employer, and SDX Alliance may not be able to use it. If they can't, they will need to rewrite the entire code, delaying the release of the product. & What should Dona's course of action be if it is found that they are not legally allowed to use the code? & Leave it in and hope no one notices & Wrong & Demonstrate Integrity (EA), Judgement (ACM) \\

& & Leave it up to managers to decide & Not desirable & \\
& & Perform minor rewrites of sections of the code to try and obfuscate the issue & Bearable & \\
& & Perform a full rewrite of the code and delay the release of the product & Ideal & \\

\cline{2-5}
& Who was most responsible for knowing whether Dona's code could be reused when developing for SDX Alliance? & Dona's previous employer & Wrong & Practice Competently (EA), Competence (ACS) \\

& & Dona's manager & Not desirable & \\
& & SDX Alliance's lawyers & Bearable & \\
& & Dona & Ideal & \\
\cline{2-5}

& If you were the employee in the quality assurance team of SDX Alliance who discovered the copyright issue, what would you do? & Keep it to yourself and allow product release & Wrong & Exercise Leadership (EA) \\
& & Suggest to Dona to modify code so that it is different enough from the old one & Not desirable & \\
& & Privately bring the issue up with Dona and let them decide & Bearable & \\
& & Report the issue despite the potential delay it will cause & Ideal & \\
\hline
\end{tabular}
\end{table*}
The key challenge that we faced in designing the Software Ethics Quiz was to draft the quiz contents including scenarios, questions, and answer options. To begin, we brainstormed possible scenarios and questions for our quiz. Our aim was to develop a quiz that contained scenario-based questions that posed hypothetical ethical dilemmas and quiz-takers would be tasked with choosing the most suitable (aka ideal) responses aligned with the ethical principles they had been exposed to in the lecture. Scenario-based questions are one of the most effective questions to ask as they foster engagement and can help stimulate a person’s thought and decision-making process \cite{lally2017watsonpaths}. The quiz comprised a total of three scenarios, with three questions accompanying each scenario. Each question offered four answer choices, and participants had to choose their preferred course of action when facing that scenario. 

All three scenarios were designed based on inspirations from real-world software ethical issues and incidents while aligning their spread across the codes of ethics. For example, the first scenario that we designed for our quiz was inspired by the \emph{Optus 2022 Data Breach} incident that affected millions of users in Australia and highlighted the importance of the security of user data and the impacts of unethical software \cite{Optus}. The second scenario in our quiz took inspiration from various incidents arising from the incentives of social media companies, resulting in the formation of echo chambers. \cite{Socialmedia,cinelli2021echo}. The third scenario of our quiz was inspired by a case study regarding copyright concerns where an entry-level software engineer's code was found copyrighted during the quality assurance process \cite{Copyright}. The details of the scenarios of the quiz can be found in Table \ref{tab:MappingTable}.

\begin{figure}
    \centering
    \includegraphics[width=\columnwidth] {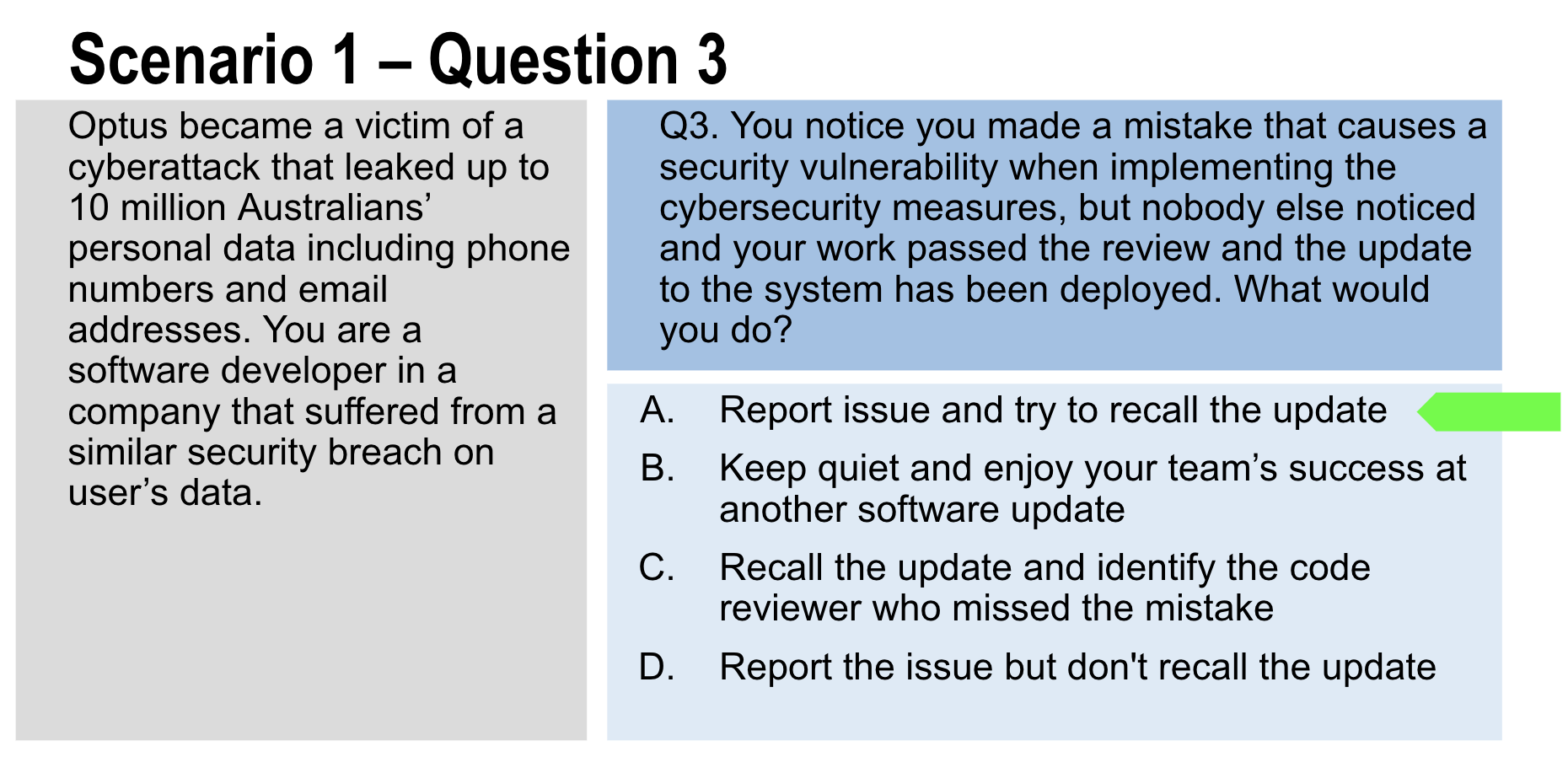}
    \caption{Scenario 1, Question 3 along with its answer options and the ideal answer with an animated green tag}
    \label{fig:Summary}
\end{figure}

While designing the scenarios and quiz questions, we ensured that the language used was as neutral, unbiased, and non-leading as possible. This included ensuring similar length, language, and complexity of all answer options. For example, Figure \ref{fig:Summary} shows one of the scenarios and its related question-and-answer options, with the ideal answer highlighted with an animated green tag. The full list of scenarios and respective questions and answers, evaluation of answers, and the applicable code of ethics can be seen in Table \ref{tab:MappingTable}. 

All the team members (four research students and three research supervisors) were involved in brainstorming the scenarios, and developing the quiz questions and answer options. Since the topic of \emph{ethics} can often be subjective and grey, different people may have diverse interpretations of this topic. Consequently, we engaged in several rounds of team discussions and brainstorming to design the scenarios, questions, and answer options for the quiz. We had open discussions among the team members on the aspects we did not agree on. We shared different ideas and found common ground on each other's perspectives while always grounding the discussion in the selected code of ethics.

The ACM/IEEE Software Code of Ethics \cite{IEEECS} was considered as it applies to the wider international SE community. We considered the Australian Computer Society Code of Ethics \cite{ACS} and the Engineers Australia Code of Ethics \cite{EngineersAustralia} in the design of the quiz as (i) these are well-recognised codes of ethics in the software industry, (ii) our primary objective was to design the quiz for the  SE students studying and working in Australia, and (iii) the Bachelor of Software Engineering degree they were enrolled in is accredited by the ACS and EA and so adhering to their code of ethics was a professional requirement. Our aim was to encompass most of the principles of the ACM/IEEE Software Code of Ethics \cite{IEEECS}, the ACS Code of Ethics \cite{ACS}, and the EA Code of Ethics \cite{EngineersAustralia} by consolidating duplicated and similar codes of ethics. The quiz structure involves presenting each scenario rooted in a principle from either the ACM/IEEE Software Code of Ethics, the ACS Code of Ethics, or the EA Code of Ethics, followed by three 
questions derived from principles of either code of ethics. The integration of the software code of ethics not only reinforces the quiz's theoretical foundation but also fulfills its educational objectives.

\subsubsection{Scoring and Feedback}
Another challenge that we faced in the design of our Software Ethics Quiz was to develop a meaningful scoring and feedback system. Our aim was to provide overall feedback and scores to the SE students on their answer choices so they could learn about software ethics while taking the quiz. 

We formulated the answer choices in a manner that did not simply indicate correct or incorrect responses. After multiple rounds of discussions and reviews, we decided to design our answer options as `wrong', `not desirable', `bearable', and `ideal' with a  score of 1, 2, 3, and 4 respectively. A score of 1 was given to the wrong answer as a means of appreciating the effort made in engaging with the material and to encourage responses. The quiz concludes by summing up the scores from individual questions to provide a total percentage. Additionally, it provides feedback to students based on their scores. Table \ref{tab:Feedback} shows the overall feedback provided to users based on their total scores. 

\begin{table}[]
\caption{Scores and Feedback of the Quiz} 
\begin{center}
\begin{tabular}{|p{1.1cm}|p{6.5cm}|}
\hline
\textbf{Score} & \textbf{Feedback} \\
\hline
76-100\% & Excellent attempt. Congratulations! You are a software ethics expert.\\
\hline
51-75\% & Very good attempt. You are on track to becoming a software ethics expert.\\
\hline
26-50\% & Satisfactory attempt. With a little more knowledge, you can become a software ethics expert.\\
\hline
 0-25\% & Unsatisfactory attempt. There is much room for improvement.\\
\hline
\end{tabular}
\label{tab:Feedback}
\end{center}
\end{table}

\subsection{Software Ethics Quiz -- Use and Evaluation}

\subsubsection{Ethical research considerations}
Since one of the authors was the instructor, it was important to follow a clear separation of concerns when delivering the quiz. To this end, as per the requirements of the Human Research Ethics Committee (HREC) at \textbf{anonymous} University (approval number suppressed for anonymity), the students were informed about the research in advance, including being shared the explanatory statement and consent form on the online learning management system. It included highlighting the anonymous nature of the quiz, the students' discretion in having their anonymous data included in the research and the voluntary and anonymous nature of the feedback survey. These are standard measures of ethical research employed in educational research to ensure students' privacy, learning experience, and psychological safety through careful consideration of avoiding any real or perceived sense of the impact of the research on students' learning experience and/or grades. For example, by separating the feedback survey from the quiz, we enabled all students to equally participate in the learning opportunity offered by the quiz while only those interested in sharing their feedback filled the post-quiz survey. All aspects of the quiz and feedback were kept anonymous.

\subsubsection{Class context}
After obtaining ethics approval from the HREC, we decided to use the Software Ethics Quiz as an educational resource with the students undertaking a second-year SE class at \textbf{anonymous University} across its \textbf{anonymous campuses}. This class was a core requirement for the students enrolled in the Bachelor of Software Engineering and an elective (or optional) class for those enrolled in similar degree programs such as the Bachelor of Computer Science and double degrees. One of the key learning outcomes of the unit was focused exclusively on ethical considerations in SE and applying relevant codes of ethics. The ethics content was covered in Week 3 of the semester which was twelve weeks long.

\subsubsection{Ethics coverage}
The coverage of ethics in the class was structured as follows:\\

\noindent \textbf{Ethics Lecture:} A one-hour online lecture was delivered on Tuesday of the week which covered key ethical concepts including definitions as shared earlier; differences between ethics, morality, and law; motivating the need and importance of ethics in SE; introduction to the three most relevant code of ethics including the ACM/IEEE Software
Code of Ethics \cite{IEEECS}, the ACS Code of Ethics \cite{ACS}, and the EA Code
of Ethics \cite{EngineersAustralia} for the students as explained earlier; and a brief overview of related aspects such as intellectual property, copyright, software licensing, open source software, and software liability. The lecture was delivered live to those attending and the recording was shared on the learning management system for those viewing it in their own time. Links to the relevant codes of ethics were also shared with students as educational resources.\\
    
\noindent \textbf{Ethics Workshop:} Five two-hour in-person workshops were conducted on Thursday and Friday across the two campuses. The first hour of these workshops was dedicated to ethics and the remaining time to other class activities. The workshop included a very brief overview and reminder of the contents of the lecture to bring everyone on the same page, using a set of slides displayed in the workshop. This was followed by sharing a link and a QR code for the students to access and take the Quiz (on Google Form) anonymously. 

\noindent \faQuestionCircle[regular] ~\textit{Individual Student Quiz:} Students were instructed to complete the quiz independently, refraining from discussing it with their peers to ensure that each student's answers remained uninfluenced by others. The presence of teaching assistants around the room ensured this was followed. The instructor reminded the students that a whole class discussion would be conducted once everyone was done. They were also reminded that it was anonymous and that there were no rewards or marks associated with the quiz. Students were given approximately 10 minutes to complete the quiz. Additional time was given to those still finishing until all students present in the workshops had taken the quiz. Students could see their overall percentage score and the associated feedback (see Table \ref{tab:Feedback}), but could not see how well they performed on individual questions. This was done purposefully, so as to reveal the ideal responses and discuss the other options collectively during the debrief session.\\

\noindent \faUsers~\textit{Collaborative Debrief:} Once all students had submitted their individual feedback, the next set of slides was presented. These included each scenario and question on a single slide along with its four possible answer options, as depicted in Figure \ref{fig:Summary} for Scenario 1 -- question 3. At this point, the instructor asked students to discuss their answers with their team members. Students could do this easily as they were sitting, as usual, with their project team members (n=5 or 6). They were given a minute or two to do this. Then they were asked to share their thoughts as to the ideal answer and their justification for it. The teaching assistants moved around with the mics to capture students' responses. Differences in opinions were openly discussed, following a reminder of psychological safety and no judging of answers. Once students had shared their thoughts and discussed -- and in some cases actively debated the options -- the instructor revealed the ideal response (animated green tag) and discussed why that was the ideal response given the scenario, with a reference to the relevant codes of ethics. Then they went through all the other answers in turn, discussing why they were desirable (but not ideal), bearable, and wrong, in each case. This sparked additional questioning from the students and made for an engaging discussion.\\
        
\noindent \faPoll~\textit{Individual Student Feedback:} The last step was to collect anonymous feedback from the students during the workshop. Same as before, a QR code was displayed on the screen that the students could scan to access the feedback form. A time frame of approximately five minutes was allocated to complete it. The students were made aware that providing feedback was optional and could choose not to provide their anonymous feedback without any impact on their experience or grades. The feedback mainly focused on the interactive approach to teaching ethics to help us understand what worked well, and what did not, and how we can improve these resources for teaching ethics in the future. In this sense, the project was simply collecting feedback on our attempts to teach ethics in a new and more engaging way in the class. We wanted to use the feedback to improve future delivery and also share our experiences with the wider international education research community as we believe that if educators do not share their teaching innovations and experiences, improvements will remain localised.

The content covered through the lecture and workshop was assessable in the upcoming test (25\% worth) in Week 4. 

\section{Findings} \label{sec:findings}

\subsection{SE Quiz Responses}
Our first research question (RQ1) was: \emph{To what extent do software engineering students make ethical decisions when faced with hypothetical scenarios?} 

The aim of the quiz was to evaluate the software ethics literacy of the participants. We received a total of 347 responses to the quiz, concluding the quiz with a mandatory question seeking permission to incorporate their responses for research. Out of these, 100 students chose not to have their anonymous responses included in the research, and one response was incomplete, so we had 246 usable responses that we could analyse and report on. 

\begin{figure}[h]
    \centering
    \includegraphics[width= \columnwidth] {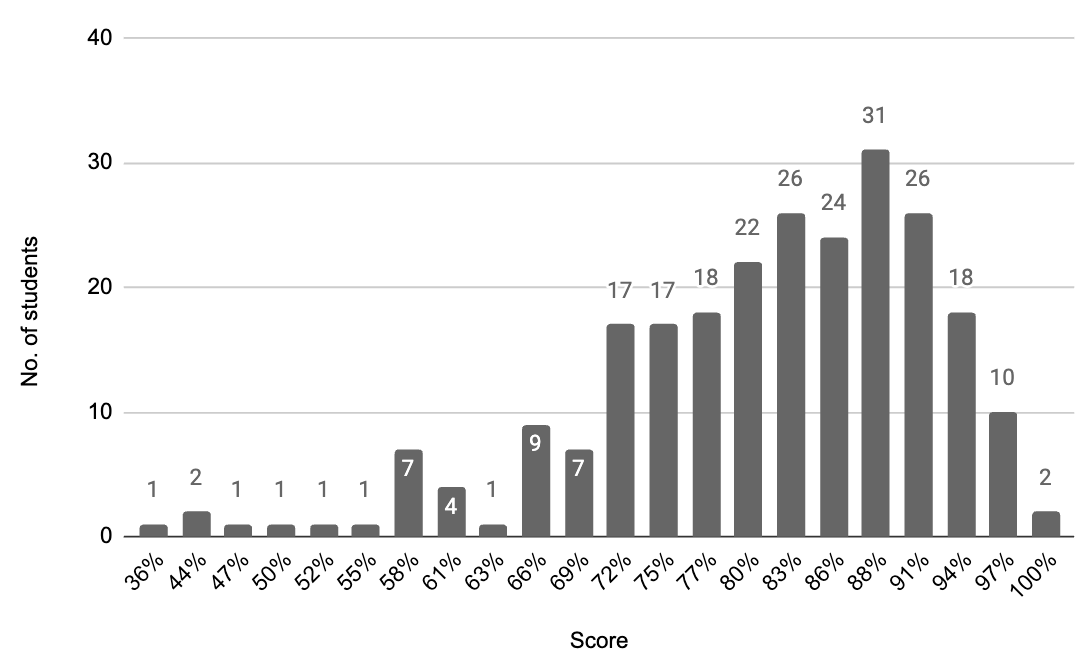}
    \caption{Distribution of scores across the 246 respondents }
    \label{fig:Score Distribution}
\end{figure}

The mean quiz score for answering all nine questions was 29.35 out of 36 (81.50\%). The maximum score attained was 36 (100\%), attained by two students. The minimum score attained was 13 (36\%), attained by one student. It can be inferred that the sample of willing research participants was fairly competent in their understanding of software ethics. Figure \ref{fig:Score Distribution} shows the distribution of scores from 36\% to 100\% across the 246 respondents. 

\subsection{Student Feedback on SE Quiz Responses} 
The objective of using the feedback form was to evaluate the efficacy and engagement of teaching and learning about ethics in the online (live and recorded) lecture using slide content on ethics and in the in-person workshop using the quiz and the associated debriefing session. There were 147 anonymous responses in total for the voluntary feedback form. Seven participants stated that they had not taken part in the lecture. Since we aimed to compare the quiz with the lecture, we did not include these responses. Three exhibited inconsistency in answers that showed discrepancies in whether or not they had attended lectures and the workshop, so we excluded those as well. Overall, we had 137 valid and usable responses to the student feedback form.

The feedback form aimed at evaluating the lecture and the quiz, and contained eight questions. Six questions provided a Likert-Scale (1 = strongly disagree, 5 = strongly agree, Did not attend/take),  and two questions provided space for open answers. The findings report on the six questions containing predefined answers using descriptive statistics. The discussion is complemented by findings from the two open questions. A full list of the questions in the feedback form along with answer options is included in Table \ref{tab:FeedbackQuestions}.

\begin{table*}
\centering
\caption{A full list of the questions in the ethics in SE feedback form}
\label{tab:FeedbackQuestions}
\footnotesize
\begin{tabular} {>{\raggedright\arraybackslash}p{0.4cm}>{\raggedright\arraybackslash}p{7cm}>{\raggedright\arraybackslash}p{7cm}} 

\hline

S.N. & Question in the Feedback Form & Answer Options \\
\hline

1 & [Lecture] Learning about the different codes of ethics (e.g. ACS/EA codes of ethics) in the lecture helped improve my understanding of ethics in software engineering & 1-Strongly Disagree, 2-Somewhat Disagree, 3- Neither Agree nor Disagree, 4-Somewhat Agree, 5-Strongly Agree, Didn't attend/take  \\

2 & [Lecture] The lecture was an engaging way to learn about ethics in software engineering & 1-Strongly Disagree, 2-Somewhat Disagree, 3- Neither Agree nor Disagree, 4-Somewhat Agree, 5-Strongly Agree, Didn't attend/take  \\

3 & [Workshop] Taking the ethics in SE quiz in the workshop helped improve my understanding of ethics in software engineering & 1-Strongly Disagree, 2-Somewhat Disagree, 3- Neither Agree nor Disagree, 4-Somewhat Agree, 5-Strongly Agree, Didn't attend/take  \\

4 & [Workshop] The ethics in SE quiz was an engaging way to learn about ethics in software engineering & 1-Strongly Disagree, 2-Somewhat Disagree, 3- Neither Agree nor Disagree, 4-Somewhat Agree, 5-Strongly Agree, Didn't attend/take  \\

5 &[Workshop] I found it hard to identify the most ideal responses to the questions in the ethics in SE quiz & 1-Strongly Disagree, 2-Somewhat Disagree, 3- Neither Agree nor Disagree, 4-Somewhat Agree, 5-Strongly Agree, Didn't attend/take  \\

6 & [Workshop] I found the debrief discussion on the scenarios and responses of the ethics in SE quiz helped me learn about ethics in software engineering & 1-Strongly Disagree, 2-Somewhat Disagree, 3- Neither Agree nor Disagree, 4-Somewhat Agree, 5-Strongly Agree, Didn't attend/take  \\
7 & What did you like the most about today’s workshop? & Open-text answer \\
8 & Please share some areas of improvement that we can apply. & Open-text answer\\
\hline

\end{tabular}
\end{table*}

\subsubsection{Exploring Efficacy}

Our second research question (RQ2) was: \textit{How helpful is an online lecture in improving students' awareness and understanding of ethics? How helpful is an interactive in-person quiz for the same?}

\begin{figure}
    \centering
    \includegraphics[width=\columnwidth] {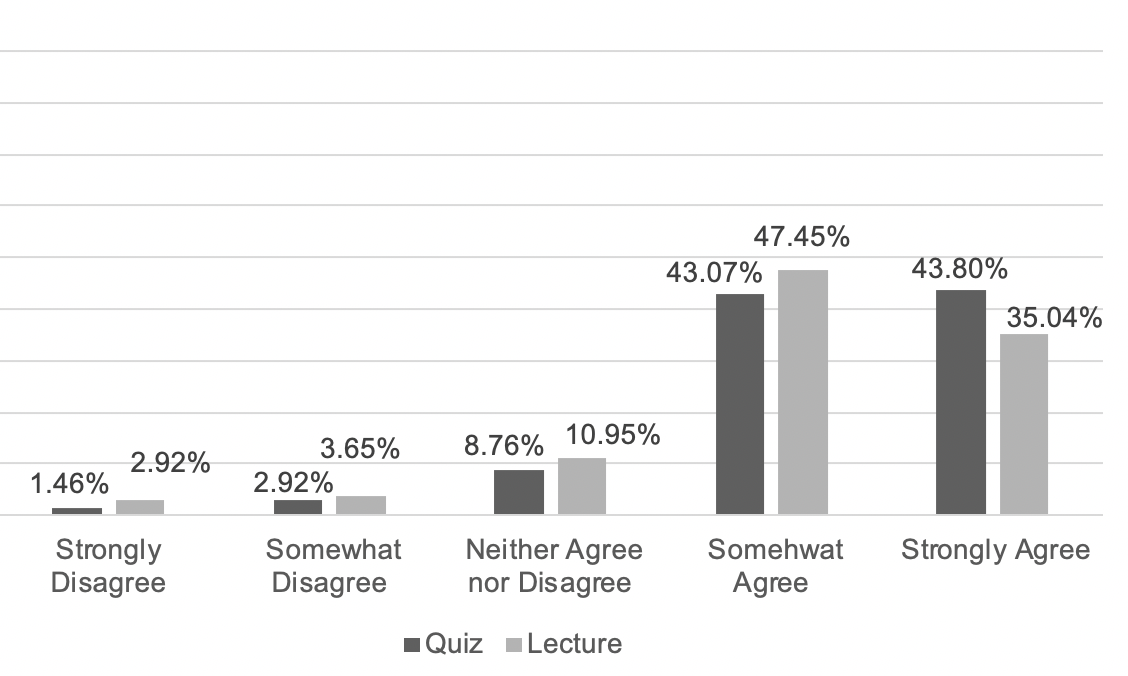 }
    \caption{Student responses about the efficacy of the quiz and the lecture}
    \label{fig:Lecture Efficacy}
\end{figure}

A majority (86.87\%) of the students agreed or strongly agreed that the quiz helped them improve their understanding of ethics in SE as shown in Figure \ref{fig:Lecture Efficacy}. While 8.76\% neither agreed nor disagreed, and 4.38\% disagreed or strongly disagreed that the quiz helped them improve their understanding of ethics
in SE. Most (82.48\%) of students agreed or strongly agreed that the preceding lecture about software engineering codes of ethics helped them improve their understanding as shown in Figure \ref{fig:Lecture Efficacy}, whereas 10.95\% neither agreed nor disagreed, and 6.57\% disagreed or strongly disagreed that the preceding lecture about software engineering codes of ethics helped them improve
their understanding. Although students were not asked to explicitly compare the two, the perceived improvement in understanding of ethics in SE was slightly higher (4.39 percentage points) for the quiz than for the lecture.

\begin{figure}
    \centering
    \includegraphics[width=\columnwidth] {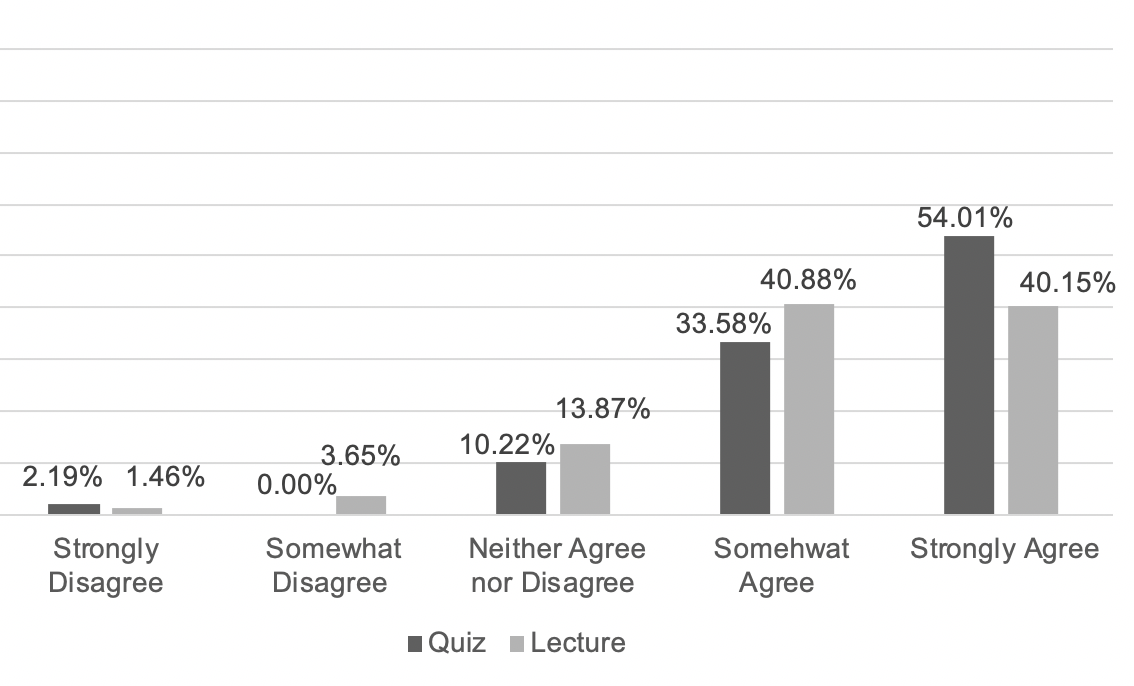 }
    \caption{Student responses about how engaging the quiz and the lecture were}
    \label{fig:Lecture Engagement}
\end{figure}

\subsubsection{Exploring Engagement} 

Our third research question (RQ3) was: \textit{How engaging is an online lecture to learn about ethics in software engineering? How engaging is an interactive in-person quiz for the same?}

A majority (87.59\%) of students found that the quiz was an engaging way to learn about ethics in SE as shown in Figure \ref{fig:Lecture Engagement}. 10.22\% neither agreed nor disagreed, and 2.19\% strongly disagreed that the quiz was an engaging way to learn about ethics in SE. This is in some contrast to the 81.02\% of students who felt the same about the lecture as shown in Figure \ref{fig:Lecture Engagement}. Few students (13.87\%) neither agreed nor disagreed that the lecture was engaging, and 5.11\% disagreed or strongly disagreed. Similar to RQ2, the difference between the perceived engagement of the quiz versus that of the lecture was 6.57\% in favor of the quiz for RQ3.

\subsubsection{Difficulty of Answering the Quiz} 
Our fourth research question (RQ4) was: \textit{How difficult is it to find the most ideal answers to the quiz questions?}

\begin{figure}
    \centering
    \includegraphics[width= \columnwidth] {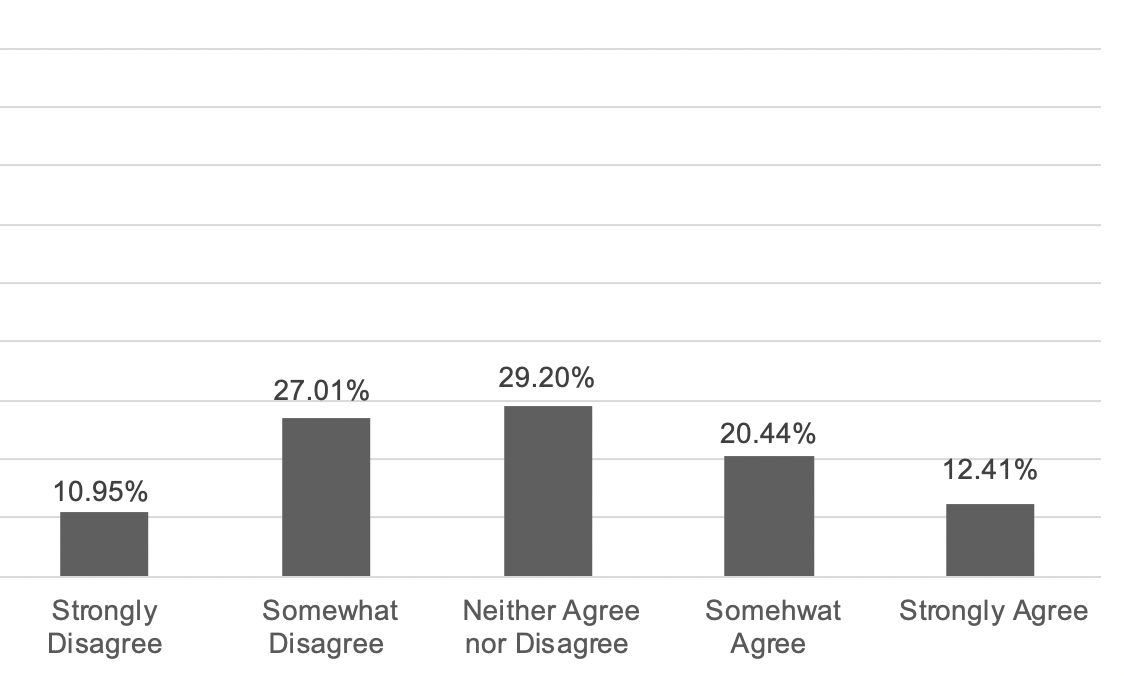}
    \caption{Student responses about the difficulty of identifying ideal answers in the quiz}
    \label{fig:Ideal responses}
\end{figure}

The students' perception of the difficulty of identifying ideal answers in the quiz varied. For example, 32.85\% of students agreed or strongly agreed that it was hard to identify the ideal responses to the questions in the \emph{Software Ethics Quiz}, while 37.96\% of students disagreed or strongly disagreed that it was hard to identify the ideal responses to the questions in the \emph{Software Ethics Quiz}. Some 29.20\% of students neither agreed nor disagreed that it was hard to identify the ideal responses to the questions in the \emph{Software Ethics Quiz} as shown in Figure \ref{fig:Ideal responses}. This helped answer the fourth research question (RQ4) and revealed that the perceived difficulty of finding the most ideal answers to the quiz questions varies by individual.

\subsubsection{Effectiveness of the Debrief Session} 
Our fifth research question (RQ5) was: \emph{How helpful is a debrief discussion on the quiz's scenarios and student responses to learn about ethics in software engineering?}

\begin{figure}
    \centering
    \includegraphics[width= \columnwidth] {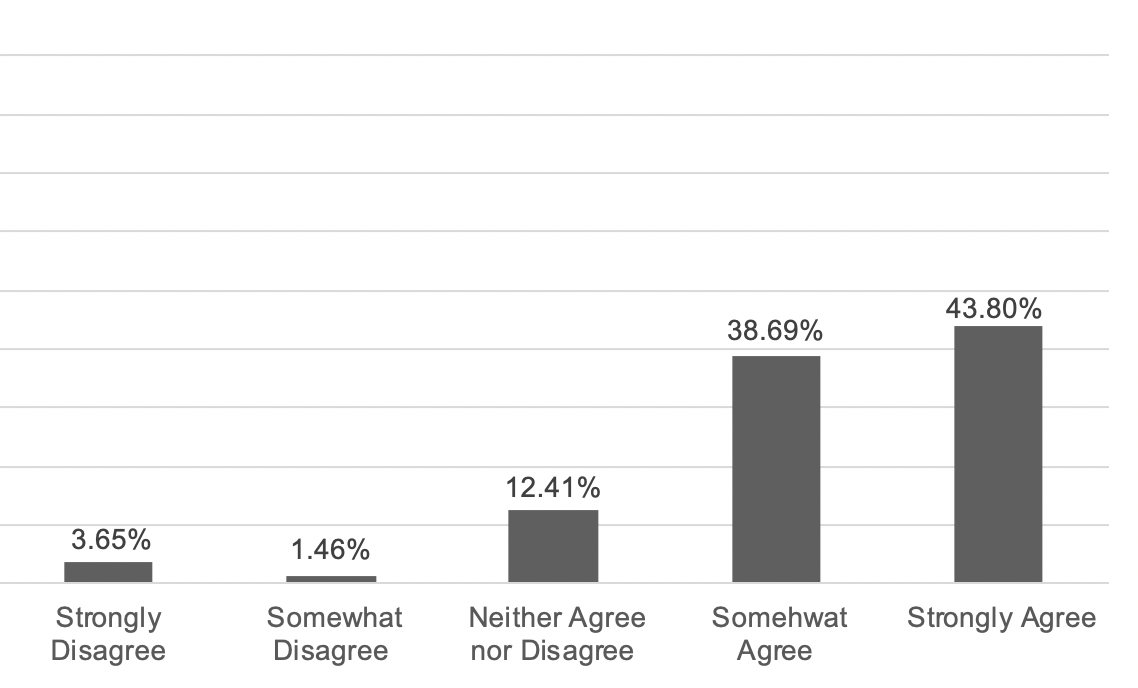}
    \caption{Student responses about the effectiveness of a debrief discussion}
    \label{fig:Debrief discussion}
\end{figure}

A majority of the students found the debrief session during the workshop very helpful in learning about ethics in SE. A significant 82.48\% of the participants either strongly agreed or somewhat agreed when asked if the session was beneficial for gaining insights into ethics
in SE, whereas only 5.11\% of the participants disagreed as shown in Figure \ref{fig:Debrief discussion}. Likewise, 12.41\% of the participants remained neutral when asked about the effectiveness of the debrief session. This helped answer the fifth research question (RQ5), i.e. a majority of the participants found the debrief discussion effective for learning and obtaining insights on the ethics in SE.

\section{Discussion}

Our study aimed at understanding how we can teach the next generation of software engineers about software ethics so they are better aware of ethical codes and guidelines and how can we do this in ways that are engaging. In this section, we provide a summary of the quantitative findings presented earlier; a summary that incorporates some insights obtained from the participants' open-ended responses to the feedback survey (qualitative findings); limitations of our study and approach; and a list of implications including ideas for future work.


\subsection{Summary of Quantitative Findings}
Our results indicate that the SE students enrolled in our class had gained a respectable understanding of software ethics with a mean value of 29.35 out of 36 (81.5\%). Two students achieved a perfect score of 36 out of 36, while the lowest score attained was 13 by one student. Based on descriptive statistics, we find that overall, the students felt that both the lecture and the quiz helped improve their understanding of software ethics. The perceived improvement in understanding ethics in SE was slightly higher (4.39 percentage points) for the quiz than for the lecture. At first glance, this may seem to suggest that a scenario-based quiz is more useful than a regular lecture. However, the role of the lecture in establishing awareness and understanding of the relevant codes of ethics ahead of the quiz cannot be discounted. As such, we interpret these results to be supportive of a combined approach. Similarly, participants rated engagement in the quiz as 6.57 percentage points higher than that for the lecture. This points in the direction that students felt more engaged in the quiz than attending the lecture. The fact that the lecture was delivered online and seen by students either live or recorded and the quiz was conducted interactively in person can help explain this difference.

The answers regarding the difficulty of the quiz varied among participants and were more or less equally distributed. Therefore, it can be said that the difficulty of finding the most ideal answers to the questions of the quiz depends on the individual. It can also be inferred that the team's effort in designing the quiz to be sufficiently challenging, but not overwhelming, had worked to an extent.

The debrief discussion on the quiz scenarios and responses was seen as highly valuable by participants in gaining deeper insights and comprehension of ethics in SE. This suggests that incorporating debrief sessions after taking the quiz could potentially enhance the learning experience for students studying ethics in SE. It will also serve to emphasize the importance of open dialogues within software development teams that the students can bring to their future practice.

Overall, based on the results of our small-scale evaluation in a class, the scenario-based quiz along with a debrief session shows potential to be an engaging approach to raise awareness of software ethics among future software engineers.

\subsection{Summary of Qualitative Findings}

In addition to the multiple-choice questions, the feedback form included two open-ended questions. These were:

\begin{itemize}
    \item \textbf{Open-ended question 1:} What did you like the most about today’s workshop?
    \item \textbf{Open-ended question 2:} Please share some areas of improvement that we can apply.
\end{itemize}

We received short comments in response to these questions. In response to the first open-ended question, most students mentioned the \textit{quiz} and how \textit{engaging} it was. One participant noted \faComment~``\textit{The quiz was an interactive way to get people thinking about ethics more practically}'' while another said, \faComment~``\textit{Having a Quiz is better than just reading slides}''. Some liked the \faComment~``\textit{real life scenarios}''.  The interactive nature of the debrief session was also mentioned, e.g. \faComment~``\textit{Student voicing their reasoning for answers}'' and 
\faComment~``\textit{When people argued for their opinions. I thought it was great and engaging.}''

In response to the second open-ended question, some ideas mentioned included allowing students to review their individual answers and using a more interactive format such as Flux or Kahoot, \faComment~``\textit{More interactive format like Kahoot maybe}''. In other feedback, some students asked for more time to discuss the scenarios and questions while others thought it could have been shorter. Based on this feedback, we share our ideas for future work in the last section.

\subsection{Limitations and Threats to Validity}
We now discuss the limitations of our study and approach. While a majority of the students enrolled in the class participated and agreed to have their anonymous responses used for research purposes, we cannot claim wider generalisability to a larger SE student population since our findings are limited to one university and class. Designing the scenarios so they were grounded in reality while also addressing key codes of ethics was non-trivial. Keeping the workshop length under reasonable limits was also a key concern to maintain student engagement, make the lesson practically workable as a classroom, and leave enough time for the debriefing. As a result, the scenarios and questions do not map to each and every code of ethics from the three relevant sources, rather, they cover some of them.  The questions of our quiz cover the following ethical codes:  \emph{Public} and \emph{Judgement} from ACM/IEEE Software
Code of Ethics \cite{IEEECS}, \emph{Demonstrate Integrity}, \emph{Promote Sustainability}, \emph{Exercise Leadership} and \emph{Practice Competently} from  EA Code
of Ethics \cite{ACS}, and \emph{The Primacy of the Public}, \emph{Honesty} and \emph{Competence} from  ACS Code of Ethics \cite{EngineersAustralia}.  

While we did not ask students to directly compare the lecture to the quiz, but rather asked them to rate them separately, their responses were slightly more appreciative of the quiz than the lecture. The fact that the quiz was taken in person with peers as compared to the lecture (live or recorded) being seen alone online may have also impacted perception.

The research team would have liked to capture other relevant data points, for example, how the students performed on the ethics questions in the upcoming test. However, to preserve students' confidentiality and psychological safety, we decided to go with an anonymous response approach. This worked well to elicit responses and permission to use for research but made it impossible for us to connect any of the responses from the quiz or feedback form to other indicators such as their test performance. This is an example of a trade-off made to sustain the comfort and trust that had been established with students in the class with our wider efforts to improve the state of ethics education through research.

\subsection{Lessons and Recommendations}

Our experience of designing ethics in SE lessons for a second-year higher education class in the Bachelor of Software Engineering program taught us some lessons that we share as recommendations for the wider education community.

\noindent \faThumbsUp~ \textbf{Use a combined approach:} Instead of selecting between a lecture or interactive approaches, our experiences lead us to recommend a combined approach. The topic of ethics includes a variety of fundamental knowledge and content that needs to be understood before useful discussions can be held. To this end, our approach of sharing the codes of ethics with the students through the university's learning management system online and covering the key concepts in an online lecture video helped students become aware of the relevant codes of ethics. This prior knowledge was the basis for the interactions we could have during the workshop when students took the quiz individually first and then had team and wider class-wide discussions about it. The combined approach can be in the form we used -- as an online lecture followed by in-person interactive workshops --- or in the form of a \textit{flipped classroom} where the lecture is pre-recorded and shared with students ahead of the workshop, as research also shows that teaching ethics solely in a classroom setting was ineffective in improving students' understanding of it \cite{bairaktarova2015engineering}.

\noindent \faThumbsUp~ \textbf{Make ethics an assessable learning outcome:} Having the ethics content assessable through a test in Week 4 may have had a positive impact on the students' engagement with the lecture and quiz. Based on our experience, we recommend simply talking about ethics may not be enough. In order to really motivate students to pay attention to the topic and to underscore its importance, ethical considerations can be added as an explicit learning outcome or as part of another learning outcome for the class. Making ethics content assessable may help elicit a good level of attention and response from students for the associated lectures, resources, and workshops.
    
\noindent \faThumbsUp~ \textbf{Acknowledge the grey:} The topic of ethics is hardly ever black and white. Our experience led us to recommend taking a balanced approach and reinforcing the idea that ethical dilemmas can be simple and straightforward sometimes, but are often more complex. This can help prepare students mentally for what is about to be discussed. We also recommend selecting or designing scenarios carefully, so there is a range of dilemmas experienced by the students, some simpler and some more complex. A similar recommendation was suggested by a study that found that a  lack of context in the questions can make the answers unreliable predictors of behavior and thereby, lead to a failure of ethics training \cite{berry2010ethics}.

\noindent \faThumbsUp~ \textbf{Establish psychological safety:} Our experiences suggest psychological safety is a key prerequisite to having useful discussions about ethics in classrooms. In our case, the instructor had established a culture of open discussions in the workshops, which helped students to share their thoughts in the ethics workshop as well. Our experiences lead us to recommend educators make conscious efforts to establish psychological safety through clear reminders of the classroom being `a safe space' and that students are not being judged and then follow through in actions. For example, encouraging a range of opinions to be voiced respectfully and reminding students that individuals can have very differing opinions when it comes to ideal ethical responses, are useful strategies. Timing the discussions after a few weeks have passed will also likely help with this as it provides time for a classroom culture and rapport between peers and between the instructors and students to be established before a topic like ethics can be discussed openly.
    
\noindent \faThumbsUp~ \textbf{Practice and teach how to talk about ethics:} Discussing ethics requires a certain level of maturity of thought and conversation: to accept differing opinions, be respectful when expressing their own opinions and when listening to others especially when they disagree. At one point, a student had become very passionate when expressing their rationales for a specific answer to a scenario. The instructor allowed the students to express themselves, and noticing the heightened emotions in the classroom, gently humored them to ``take a big breath and relax'' and the collective build-up of emotions in the classrooms could be seen to be released with laughs. Our experience of conducting the debrief session leads us to recommend educators to be prepared for passionate exchanges on the topic, and remain in control, while also injecting some humor to keep some of the heavier ethical dilemmas from becoming overwhelming for the students.

\noindent \faThumbsUp~ \textbf{Lead by example:} Students can struggle to see the practical application of ethics in everyday scenarios. After the workshop, one of the students asked the instructor how teachers practice ethics. To this, the instructor shared the example of how a human research ethics committee sits in the university and how all human-related research needs to be approved by them. They then shared concrete examples of running the ethics workshop with HREC approval and collecting anonymous student responses with consent. They also pointed to the consistent reference to third-party materials used in lectures as examples of ethical teaching practice. This was well received by the students listening. Our experiences lead us to recommend leading by example when it comes to ethics. Instructors can easily be seen as `preachy' if they do not provide examples of how they practice ethics themselves. We plan to incorporate this aspect more systematically in future offerings.

\subsection{Future Work}
Comments shared by students in their feedback form and some of the limitations gave rise to ideas for future work. As mentioned by some students in their feedback, we could have used a more interactive tool such as Kahoot but we wanted the students to take the quiz individually and without discussing or being influenced by their peers in the first instance, and then follow up with a team-based and wider classroom discussion. Google Forms worked well for this purpose. However, since we used anonymous Google forms for research purposes, it was not possible for students to review their Quiz answers. Since then we are developing a web-based version of the quiz and reviewing answers anonymously will be possible in the future. We also plan to refine the scenarios in response to how much discussion and debate they generated in the workshop in order to make some of the questions more engaging. The scenarios can also be made more or less difficult to suit students in higher classes, with more or less background knowledge of software ethics. We were able to conduct the workshop in person which had its own flavour. In the future, the same can be attempted in online settings as required while sharing experiences of how feasible an online environment is for such purposes.

Student feedback in the form of open-ended comments revealed that a few students perceived the reasoning behind the ideal answers to be inconsistent. In our next running of the class, we plan to share a more concrete and direct mapping of the justifications with the relevant codes of ethics applied. While this was done verbally, in the future, we plan to include a written explanation of the justification with a mapping to the relevant codes of ethics alongside every question.

We welcome the wider SE education community to use the quiz and feedback form -- refining them as needed to suit local conditions -- and share their experiences.

\section{Conclusion} \label{sec:conclusion}

The importance of teaching the next generation of software engineers about software ethics cannot be overstated. In this paper, we shared our experiences of teaching software ethics to a second-year Bachelor of Software Engineering class in a higher education setting in \textbf{anonymous country}. We used a combined approach where relevant codes of ethics were shared asynchronously as \textit{resources}, the topic of ethics and the codes of ethics were covered in an online \textit{lecture} -- available live and as a recording -- and then discussed in a team-based and wider classroom-based approach following scenario-based ethics in SE quiz in the in-person \textit{workshops}. 

We developed the scenario-based \emph{Software Ethics Quiz} aligned with all three codes of ethics including the ACM/IEEE Software Code of Ethics \cite{IEEECS}, the ACS Code of Ethics \cite{ACS}, and the EA Code of Ethics \cite{EngineersAustralia}. The quiz served as a tool for SE students to self-assess their knowledge and understanding of software ethics following the lecture and resources shared. Data collected from the quiz responses and feedback form suggests a good perception of efficacy and engagement of both the lecture and quiz, with slightly better ratings for the quiz. A majority of the students found the debrief discussion on quiz scenarios very beneficial for learning about software ethics. 

We learned a number of valuable lessons from the experience that we have shared as gentle recommendations for the wider SE education community. These include: using a combined approach to teaching ethics through online lectures (or flipped classroom-type recorded lectures) and interactive workshops, making ethics an assessable learning outcome, acknowledging the complexity and multiplicity of ethics as a topic, establishing psychological safety in the classroom to encourage useful discussions, practicing and teaching how to have respectful dialogue about ethics, and leading by example through demonstrating examples of ethical behavior in teaching and learning, in other words, practice what you preach.

We hope our experiences and our quiz shared will encourage more SE educators to take on the topic of ethics in their software engineering courses and related classrooms and share their experiences so we can move forward in this critical area as a community.

\section{Data Availability}
The data that support the findings of this study are available from the corresponding author, upon reasonable request.
\bibliographystyle{ACM-Reference-Format}
\bibliography{reference}

\end{document}